\documentclass[11pt]{article}
\usepackage{moriond,epsfig}
\usepackage {color}

\def\Journal#1#2#3#4{{#1} {\bf #2}, #3 (#4)}

\def\NIMA{{\em Nucl. Instrum. Methods} A}

\def\PLB{{\em Phys. Lett.}  B}
\def\PRL{\em Phys. Rev. Lett.}
\def\PRD{{\em Phys. Rev.} D}

\def\be{\begin{equation}}
\def\ee{\end{equation}}
\def\bea{\begin{eqnarray}}
\def\eea{\end{eqnarray}}

\def\ifm#1{\relax\ifmmode#1\else$#1$\fi}
\def\DAF{DA\char8NE}  
\def\f{\ifm{\phi}}   \def\epm{\ifm{e^+e^-}}
  \def\x{\ifm{\times}}
  \def\pic{\ifm{\pi^+\pi^-}}
\def\pt#1,#2,{\ifm{#1\x10^{#2}}}  
\renewcommand{\to}{\ensuremath{\rightarrow}}

\def\pio{\ifm{\pi^0\pi^0}} 

\newcommand{\Ref}[1]{ref.~\cite{#1}}
\newcommand{\pvec}{\ensuremath{\mathbf{p}}}

\newcommand{\kl}{\ensuremath{K_L}}
\newcommand{\ks}{\ensuremath{K_S}}

\newcommand{\Ppim}{\ensuremath{\pi^-}}
\newcommand{\Ppin}{\ensuremath{\pi^0}}
\newcommand{\Ppip}{\ensuremath{\pi^+}}

\newcommand{\eV}{{e\kern-.07em V}}

\def\rmk{\rm\kern.5mm }

\newcommand{\subrm}[1]{\mbox{\tiny \rm #1}}
\begin{document}
\vspace*{4cm}
\title{CP AND CPT TESTS WITH THE KLOE DETECTOR}

\author{THE KLOE COLLABORATION\footnote {
    {F.~Ambrosino},
    {A.~Antonelli},
    {M.~Antonelli},
    {C.~Bacci},
    {P.~Beltrame},
    {G.~Bencivenni},
    {S.~Bertolucci},
    {C.~Bini},
    {C.~Bloise},
    {V.~Bocci},
    {F.~Bossi},
    {D.~Bowring},
    {P.~Branchini},
    {R.~Caloi},
    {P.~Campana},
    {G.~Capon},
    {T.~Capussela},
    {F.~Ceradini},
    {S.~Chi},
    {G.~Chiefari},
    {P.~Ciambrone},
    {S.~Conetti},
    {E.~De~Lucia},
    {A.~De~Santis},
    {P.~De~Simone},
    {G.~De~Zorzi},
    {S.~Dell'Agnello},
    {A.~Denig},
    {A.~Di~Domenico},
    {C.~Di~Donato},
    {S.~Di~Falco},
    {B.~Di~Micco},
    {A.~Doria},
    {M.~Dreucci},
    {G.~Felici},
    {A.~Ferrari},
    {M.~L.~Ferrer},
    {G.~Finocchiaro},
    {S.~Fiore},
    {C.~Forti},
    {P.~Franzini},
    {C.~Gatti},
    {P.~Gauzzi},
    {S.~Giovannella},
    {E.~Gorini},
    {E.~Graziani},
    {M.~Incagli},
    {W.~Kluge},
    {V.~Kulikov},
    {F.~Lacava},
    {G.~Lanfranchi},
    {J.~Lee-Franzini},
    {D.~Leone},
    {M.~Martini},
    {P.~Massarotti},
    {W.~Mei},
    {S.~Meola},
    {S.~Miscetti},
    {M.~Moulson},
    {S.~M\"uller},
    {F.~Murtas},
    {M.~Napolitano},
    {F.~Nguyen},
    {M.~Palutan},
    {E.~Pasqualucci},
    {A.~Passeri},
    {V.~Patera},
    {F.~Perfetto},
    {L.~Pontecorvo},
    {M.~Primavera},
    {P.~Santangelo},
    {E.~Santovetti},
    {G.~Saracino},
    {B.~Sciascia},
    {A.~Sciubba},
    {F.~Scuri},
    {I.~Sfiligoi},
    {T.~Spadaro},
    {M.~Testa},
    {L.~Tortora},
    {P.~Valente},
    {B.~Valeriani},
    {G.~Venanzoni},
    {S.~Veneziano},
    {A.~Ventura},
    {R.Versaci},
    {G.~Xu}
}}
\address{presented by MARIANNA TESTA\\Dipartimento di Fisica, Universit\`a degli studi di Roma ``La Sapienza'' and INFN Sezione di Roma, P.le A.Moro 2, 00185 - Rome, Italy }

\maketitle\abstracts{
Neutral kaons provide one of the most sensitive system to CP and CPT violation.
Tests on CP, CPT and quantum mechanics have been performed at KLOE operating at the \DAF\ \epm\ collider. Results on the quantum interference in the channel  $\f\to \ks\kl\to \Ppip\Ppim\Ppip\Ppim$, the measurement  of the BR($\kl\to \Ppip\Ppim$) and the related CP violating parameter $\epsilon$  are presented. 
Using the Bell-Steinberger relation, CPT violating parameters   have been   also obtained.}

\section{Introduction}
The KLOE detector operates at \DAF, an $\epm$ collider working at the center of mass energy $W\sim m_{\f} \sim 1.02$ GeV.   The \f\ mesons are produced essentially at rest and decay to $\ks\kl$ ($K^+K^-$) $\sim$ 34\% ($\sim$ 49\%) of the times. The $K$ mesons are produced in a pure  $J^{PC}=1^{--}$ coherent quantum state, so that observation of a \ks\ in an event signals (tags) the presence of a \kl\  and vice-versa: highly pure, almost monochromatic, back-to-back \ks\ and \kl\  beams  can be obtained.  Moreover \ks\ and \kl\ are distinguishable on the basis of their decay length: $\lambda_S \sim 0.6$ cm and $\lambda_L \sim 340 $ cm. \\
The KLOE experiment is designed to exploit the unique feature of a \f-factory environment  for the measurement of CP and CPT violation in the $K^0-\bar{K}^0$ system and more generally for the study of kaons' decays and interference.
The KLOE detector consists essentially of a drift chamber (DCH), surrounded by an electromagnetic calorimeter (EMC). A superconducting coil surrounding the barrel provides a 0.52 T magnetic field. Descriptions of the EMC and DCH can be found in \Ref{kloe:dc,kloe:emc} 


\section{Quantum interference in the channel $\kl\ks \to \Ppip\Ppim\Ppip\Ppim$.}
Test of quantum mechanics (QM) can be performed by studying the time evolution of the quantum correlated  kaon system,
in particular studying the interference pattern of the decay $\kl\ks \to \Ppip\Ppim\Ppip\Ppim$. According to QM, 
the distribution of the difference of the decay times $I(\Delta t)$ of the two kaons shows a characteristic destructive interference which prevents the two kaons from decaying into the same final state at the same time. As suggested in ref.~\cite{{bib:bertlmann},{bib:eberhard}}, a simple way to parametrize a possible   deviation of QM is to introduce a decoherence parameter $\zeta_{S,L}$ ($\zeta_{S,L}= 0$ in QM) as follows:
\begin{equation}I(|\Delta t|)\propto e^{-|\Delta t| \Gamma_L} +e^{-|\Delta t| \Gamma_S} -2\underbrace{(1-\zeta_{S,L})}_{{\mbox{\small decoherence}}} cos(\Delta m |\Delta t|) e^{-\frac{\Gamma_s+\Gamma_L}{2}|\Delta t| }
\label{eq:deltat}
\end{equation}
\begin{figure}[h]
\begin{center}
\includegraphics[width=7.0cm]{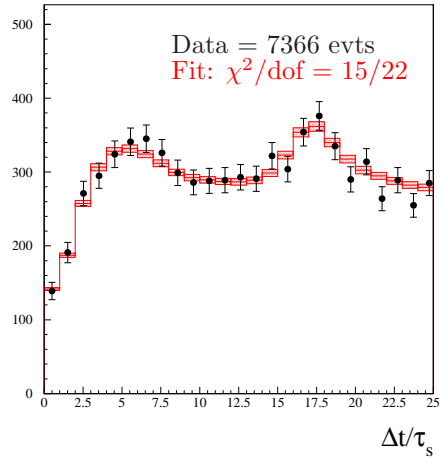}
\put(-125,155){\small Data = 7366 evts}
\put(-125,145) {\color{red}{\small Fit: $ \chi^2/$dof = 15/22}}
\caption{Fit of the difference $\Delta t$ of the decay times of $\ks\to \Ppip\Ppim$ and $\kl\to \Ppip\Ppim$. The black points are the data and the red ones are the results of the fit. The peak at $\Delta t \sim 17 \tau_S$ is due to the regeneration on the beam pipe.}
\label{fig:deltat}
\end{center}
\end{figure}
Selecting a pure sample of  $\kl\ks \to \Ppip\Ppim\Ppip\Ppim$ and fitting eq.~\ref{eq:deltat} to data, KLOE has obtained  the following preliminary result:
 $$\zeta_{S,L}=  0.043\,^{+0.038}_{{-0.035}_{\mbox{stat}}}\pm 0.008_{\mbox{syst}},$$ consistent with QM predictions. The result of the fit is shown  in fig.~\ref{fig:deltat}. 
\section{BR($\kl\to \Ppip\Ppim$)}
KLOE has measured the BR($\kl\to \Ppip\Ppim$)  using a \kl\ beam tagged by $\ks\to \Ppip\Ppim$ decays. 
The number of $\kl\to \Ppip\Ppim$
is obtained from a fit  to the $\sqrt{E^2_{miss}+|\pvec_{miss}|^2}$ 
distribution, where $E_{miss}$ is the missing energy in the hypothesis of $\kl\to \Ppip\Ppim$ decay and $\pvec_{miss}$ is the missing momentum,
with a linear combination of Monte Carlo distributions for $\kl\to \Ppip\Ppim$, $\kl\to \pi^{\pm} e^{\mp} \nu$, $\kl\to \pi^{\pm} \mu^{\mp} \nu$ 
$\kl\to \Ppip\Ppim \Ppin$ events, inclusive with respect to final-state radiation.
The number of signal events has been normalized to the number of $\kl\to \pi\mu\nu$, in order to minimize systematic uncertainties on the tagging and tracking efficiency evaluation (exploiting the similar topology of the decays as well as the momentum overlap). 
Correcting for the tagging and tracking efficiency and using the BR$(\kl\to\pi^{\pm}\mu^{\mp}\nu)$  from  ref.~\cite{KLOE:brl}, we obtain: BR$(\kl\to \Ppip\Ppim) = (1.963 \pm 0.012_{\rm stat}\pm 0.017_{\rm syst})\times 10^{-3}$. The result is in good agreement with  the measurement of KTeV~\cite{ktev} $(1.975 \pm 0.012)\times10^{-3}$ and in strong disagreement with that reported by the PDG~\cite{PDG2004}, $(2.090\pm0.025)\times10^{3}$.
This result can be used to determine  $|\eta_{+-}|$
and $|\varepsilon|$ correcting for the small contribution of $\varepsilon'$.
Using the measurements of BR($\ks\to \Ppip\Ppim$) and $\tau_{\kl}$ from KLOE
\cite{KLOE:Rs,KLOE:brl,KLOE:KLlife},  and the value of  $\tau_{\ks}$ from
PDG\cite{PDG2004} and subtracting the contribution of the photon direct emission~\cite{Dire} to the value of  BR($\kl\to \Ppip\Ppim$), we obtain:
$|\eta_{+-}| = (2.219 \pm 0.013)\times 10^{-3}$.
Finally, using the world average measurement of 
$\rm{Re}(\varepsilon'/\varepsilon)=(1.67 \pm 0.26)\times 10^{-3}$,
and assuming equal phases between $\varepsilon'$ and $\varepsilon$ we obtain 
$|\varepsilon| = (2.216 \pm 0.013)\times 10^{-3}$,
 in disagreement with the value $|\varepsilon| = (2.284 \pm 0.014)\times 10^{-3}$ 
 reported in  ref.~\cite{PDG2004}.
 The value of $|\varepsilon|$ can be 
 can be compared with the prediction~\cite{UTfit} $|\varepsilon|  = (2.875 \pm 0.455)\times 10^{-3}$
where to test the mechanism of the CP violation in the Standard Model,
 the value of $|\varepsilon|$ has been 
 computed from the measurement of the CP conserving observables:
 $\Delta {\rm m_d}$, $\Delta {\rm m_s}$,
 $V_{\rm{ub}}$, and  $V_{\rm{cb}}$.
 No significant deviation from the Standard Model prediction has been
 observed. 
\begin{figure}\label{fig:fit_klpp}
  \begin{center}
    \includegraphics[width=6.0cm]{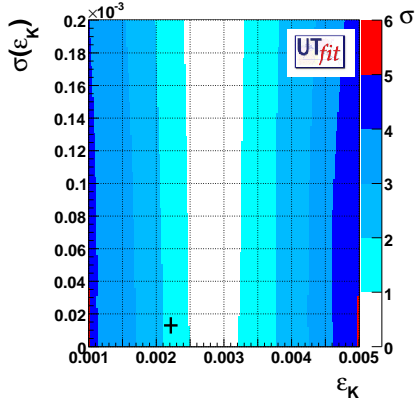}
    \caption{
$|\varepsilon|$ constraints by the  measurements 
 of $|V_{ub}|/|V_{uc}|$, $\Delta m_d$ 
 and by the limit on $\Delta m_s$on the $(\bar{\rho},\bar{\eta})$ from  ref.~\protect\cite{UTfit}, compared with the value of $|\varepsilon|$ from the measurements of BR(\kl\to\Ppip\Ppim).}
  \end{center}
\end{figure}

\section{Bell Steinberger Relation}
The most powerful test of CPT invariance in the neutral kaon system is presently obtained by means of the Bell-Steinberger relation \cite{bib:bellsteinberger}, which relates CPT and CP violating parameters, $ Im (\delta)$ and $Re(\epsilon)$, to the decay amplitudes of \kl\ and \ks\ into the same final state: 
\begin{equation}
  \begin{array}{rcl}
    (1 + i \tan{\phi_{SW}}) [ {\rm Re}(\epsilon) - i \: {\rm Im}(\delta) ]
    = {\displaystyle{\sum_{{\subrm{final}} \atop {\subrm{states}} \: f}}} A(\kl \to f)^\star A(\ks \to f) / \Gamma_S 
     =  {\displaystyle{\sum_{{\subrm{final}} \atop {\subrm{states}} \: f}}} \alpha_f
  \end{array}
   \label{eqn:cpt1}
\end{equation}
where $\phi_{SW}$ is the superweak phase, defined by $\tan{\phi_{SW}} = 2\Delta M/(\Gamma_S-\Gamma_L)$. 
For the determination of the $\alpha_f$ parameters,
experimental inputs are \kl\ and \ks\ branching
ratios, the relative phases between the amplitudes, and the \kl\ and \ks\
lifetimes, $\tau_{\ks}$ and $\tau_{\kl}$. 
We use the value of 
$\tau_{\ks}$ reported by the PDG~\cite{PDG2004} and  $\tau_{\kl}$ from KLOE average~\cite{KLOE:brl,KLOE:KLlife}
and the following measurements:
\begin{itemize}
 \item  the new KLOE  measurement of  
        BR(\ks\to\Ppip\Ppim), BR(\ks\to\Ppin\Ppin) from ref~\cite{KLOE:Rs}
	which enters in the evaluation of  $|\alpha_{+-}|$ and  $|\alpha_{00}|$,
 \item  the average between the  BR(\kl\to\Ppip\Ppim)
        here presented and that measured by  KTeV\cite{ktev}, used to  determine  $|\alpha_{+-}|$,
 \item  the measurement of BR(\kl\to\Ppin\Ppin) from KTeV \cite{ktev},
 used to  determine       $|\alpha_{00}|$,
\item  the values, $\phi_{+-}$ and $\phi_{00}$,  of the phases of $\alpha_{+-}$
  and $\alpha_{00}$, taken from the PDG\cite{PDG2004} fit without assuming CPT symmetry,
\item the measurement of the CP conserving
      direct component contribution 
      to the process \kl\to\pic$\gamma$ from ref~\cite{Dire} and the
      upper limit on the direct component contribution to the process
      \ks\to\pic$\gamma$ \cite{PDG2004}, both entering in the evaluation of $\alpha_{+-\gamma}$,
\item  the recent KLOE  upper limit on the BR($\ks\to \Ppin\Ppin\Ppin$)\cite{KLOE:ks3pi0}, which constraints the value of $|\alpha_{000}|$,
\item  the measurement of the BR(\ks\to \Ppip\Ppim\Ppin) reported in the
  PDG~\cite{PDG2004},    
\item the recent  KLOE measurement of the  semileptonic $K_S$ charge asymmetry $A_S$~\cite{KLOE:Rs},
  which allows  to calculate the semileptonic  contribution $\alpha_{kl3}= 2\tau_{\ks}/\tau_{\kl} B(kl3)((A_S+A_L)/4-i Im(\delta)+Im(x+))$, being $x+$ the parameter describing the $\Delta S=\Delta Q$ violation in the semileptonic decays.  $Im(x+)$ has been determined from a combined fit of $A_S$ with the semileptonic time dependent decay rate asymmetry measured by CPLEAR~\cite{cplear2}. The semileptonic $K_L$ charge asymmetry $A_L$ has been taken from the PDG~\cite{PDG2004}.
\end{itemize}                           
As $\phi_{+-\gamma}$, the phase of $\alpha_{+-\gamma}$, has not been measured yet, no constraints have been assumed on its value. Using these experimental inputs, the differences between  the \ks\ and \kl\ masses and lifetimes, $\Delta M$ and $\Delta \Gamma$, reported in the PDG\cite{PDG2004} (for the determination of $\phi_{SW}$), we obtain:
${\rm Re}(\epsilon) = (160.2\pm 1.3)\times 10^{-5}$ and ${\rm Im}(\delta) = (1.2\pm 1.9)\times 10^{-5}$, resulting in a considerable improvement to the CPLEAR measurement~\cite{cplear}: ${\rm Re}(\epsilon) = (164.9\pm 2.5)\times 10^{-5}$ and ${\rm Im}(\delta) = (2.4\pm 5.0)\times 10^{-5}$.

\end{document}
